\documentclass[a4paper,fleqn]{cas-dc}

\usepackage[numbers]{natbib}

\usepackage{amsmath,amsfonts}
\usepackage{algorithmic}
\usepackage{graphicx}
\usepackage{textcomp}
\usepackage{todonotes}
\usepackage[ruled,vlined,linesnumbered]{algorithm2e}
\usepackage{soul}
\usepackage{svg}
\usepackage{caption}
\usepackage{amsmath}
\usepackage{booktabs}
\usepackage{hyperref}
\usepackage{balance}
\usepackage{tabulary}
\usepackage{multicol}
\usepackage{stfloats}
\usepackage{multirow}
\usepackage{subcaption}

\usepackage{mathtools, cuted}

\newcommand{\subsubsubsection}[1]{\paragraph{#1}\mbox{}\\}

\newcommand{\sks}[1]{\textcolor{purple}{\small\textbf{[Subangkar]}#1$\triangleleft$}}

\def\tsc#1{\csdef{#1}{\textsc{\lowercase{#1}}\xspace}}
\tsc{WGM}
\tsc{QE}

\begin{document}
\let\WriteBookmarks\relax
\def\floatpagepagefraction{1}
\def\textpagefraction{.001}

\shorttitle{Simsig}    

\shortauthors{Shanto and Saha et al.}

\title [mode = title]{Contrastive Self-Supervised Learning Based Approach for Patient Similarity: A Case Study on Atrial Fibrillation Detection from PPG Signal}  



%

\author[1,2]{Subangkar Karmaker Shanto}[type=editor,
    orcid=0000-0003-0394-2565
    ]



\ead{subangkar.karmaker@gmail.com}


\credit{Conceptualization of this study, Literature review, Methodology, Machine Learning Model Design, Software, Data processing, Manuscript}

\affiliation[1]{organization={Bangladesh University of Engineering and Technology},
            city={Dhaka},
            country={Bangladesh}}

\affiliation[2]{organization={United International University},
            city={Dhaka},
            country={Bangladesh}}

\author[1,3]{Shoumik Saha}[type=editor,
    auid=000,bioid=1,
    orcid=0009-0007-7461-5306
    ]


\ead{shoumiksaha901@gmail.com}


\credit{Literature review, Methodology, Data processing, Manuscript}

\affiliation[3]{organization={University of Maryland},
            city={College Park},
            state={Maryland},
            country={United States}}

\author[1]{Atif Hasan Rahman}[type=editor,
    orcid=0000-0003-1805-3971
    ]
\ead{atif@cse.buet.ac.bd}

\credit{Problem Formulation, Supervised project, Contributed to manuscript}

\affiliation[4]{organization={United Arab Emirates University},
            city={Al Ain},
            country={United Arab Emirates}}

\author[4]{Mohammad Mehedy Masud}[type=editor,
    orcid=0000-0002-5274-5982
    ]
\ead{m.masud@uaeu.ac.ae}

\credit{Problem Formulation, Supervised project, Contributed to manuscript}

\author[1]{Mohammed Eunus Ali}[type=editor,
    orcid=0000-0002-0384-7616
    ]

\ead{eunus@cse.buet.ac.bd}

\credit{Problem Formulation, Supervised project, Contributed to manuscript}




\begin{abstract}
In this paper, we propose a novel contrastive learning based deep learning framework for patient similarity search using physiological signals. We use a  contrastive learning based approach to learn similar embeddings of patients with similar physiological signal data. We also introduce a number of neighbor selection algorithms to determine the patients with the highest similarity on the \emph{generated embeddings}. 
To validate the effectiveness of our framework for measuring patient similarity, we select the detection of Atrial Fibrillation (AF) through photoplethysmography (PPG) signals obtained from smartwatch devices as our case study. We present extensive experimentation of our framework on a dataset of over 170 individuals and compare the performance of our framework with other baseline methods on this dataset.
\end{abstract}




\maketitle

\section{Introduction}
\label{sec:introduction}

Recent advances in wearable devices and the adaptation of these devices in the medical domain for timely and continuous monitoring have led to the generation of a huge volume of patient data as a part of electronic health records (EHR). With the availability of a vast amount of data, patient similarity based diagnostics and analysis have become increasingly crucial in the medical domain. Patient similarity has been applied in several medical domains including generic diagnostics~\cite{jia2020patient, pokharel2020temporal, tashkandi2018efficient, zhang2014towards}, Alzheimer's disease~\cite{shi2022asmfs}, coronary artery disease~\cite{kolossvary2022risk}, precision medicine~\cite{brown2016patient, pai2018patient, parimbelli2018patient}, mortality prediction~\cite{lee2015personalized}, etc.

\begin{figure}
    \centering
    \captionsetup{justification=centering}
    \includegraphics[width=0.5\textwidth]{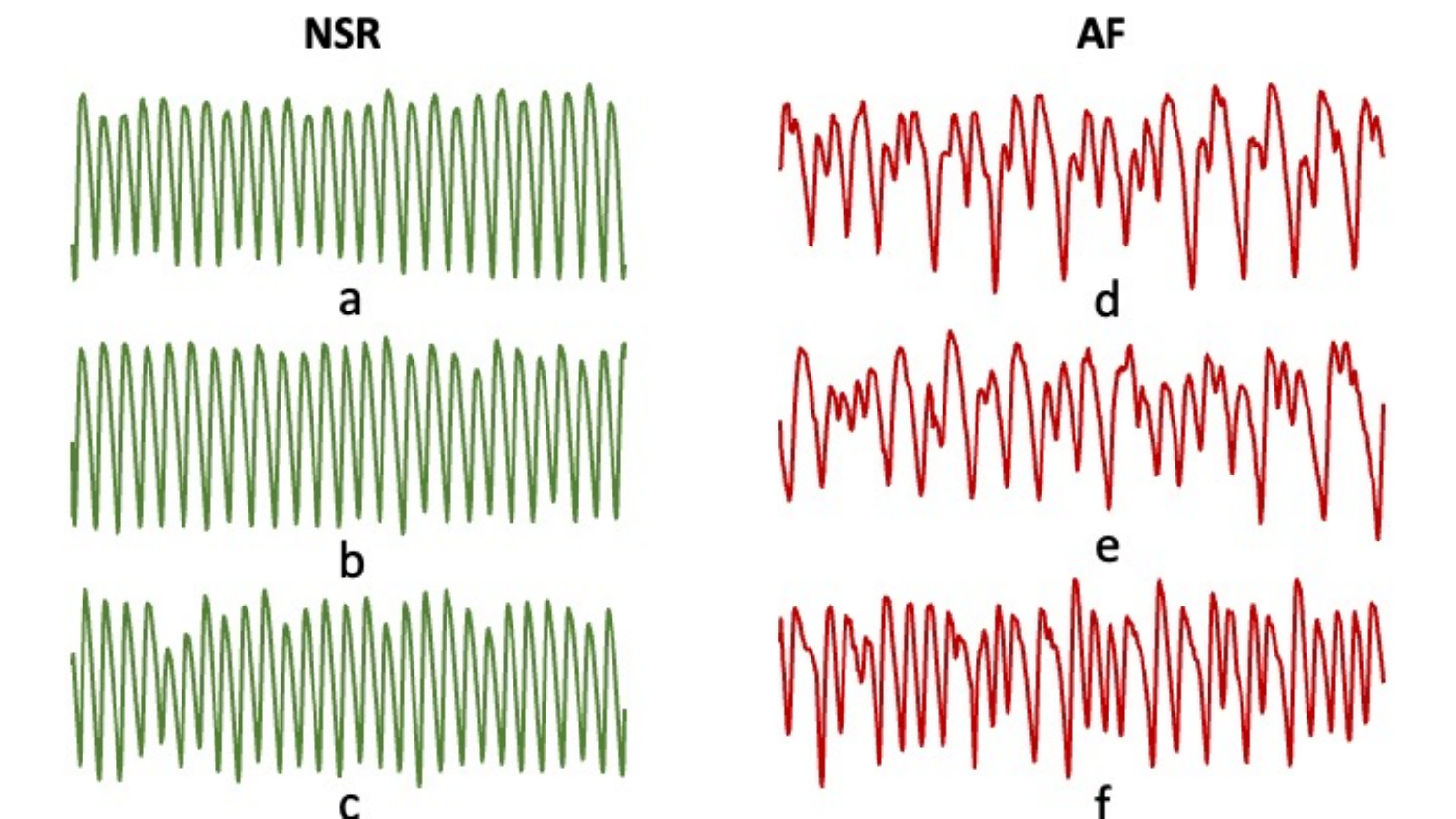}
    \caption{PPG Signals: Normal Sinus Rhythm (NSR) vs Atrial Fibrillation (AF).}
    \label{fig:NSR vs AF PPG}
\end{figure}

The key intuition of EHR based patient similarity comes from the observation that patients suffering from the same diseases or abnormalities generally preserve a common pattern. Among the sources of EHR, sensor based physiological data is one of the most common ones in recent days. For example, ECG signals are used to detect arrhythmia \& cardiovascular diseases, EEG signals are used for BCI (Brain Computer Interface), EMG signals are used for speech, prosthesis, \& rehabilitation robotics, PPG signals are used for atrial fibrillation (AF), etc.~\cite{pereira2020photoplethysmography}
Let us consider an example involving PPG signals. Figure \ref{fig:NSR vs AF PPG} shows the PPG signals for a normal sinus rhythm and atrial fibrillation. Here, varying pulse-to-pulse intervals can be noticed in the PPG signals for AF, and such abnormal patterns in signals can be utilized by similarity based learning. For example, according to Figure \ref{fig:NSR vs AF PPG}, if we want to find similar individuals as the one with signal $a$, then similarity based model will return persons with signals $b$ and $c$ as the two most similar ones. Similarly, if signal $d$ is from a query patient, then such model will return patients with signals $e$ and $f$ as the results.

Finding the similarities between pair-wise patients is one of the fundamental problems in the medical sector, and this problem has been approached by researchers from different angles over the years~\cite{allam2020patient}. 
However, most of the research relies on static data (patient background, age, weight, etc.) and longitudinal clinical events data (visit date of patients, symptoms, disease, diagnosis, etc.)~\cite{lee2015personalized, jia2020patient, zhang2014towards}, rather than physiological signals. Since incorporating time-series data like physiological signals with similarity-based learning is not trivial~\cite{m2020effective}, it has not been explored heavily in the medical domain. However, with the increasing popularity of wearable sensors in capturing physiological signals (e.g., ECG, EEG, etc.), a few recent works have also focused on signal based similarities~\cite{bianchi2018learning, wesselius2022accurate}.  The presence of noise, motion artifacts due to a lot of sensors, and missing values in some cases, still continue to pose major challenges in such research.

Researchers have also extensively studied time series data for a number of years. The recent availability of physiological signals in the medical domain has encouraged researchers to apply algorithms and models to such signals. Though statistical methods were primarily used, they suffer from low-adaptivity and robustness due to variations of signals in patients and devices. 
{They also suffer from high time complexity. One such popular algorithm for time series classification, named HIVE-COTE, has good accuracy~\cite{bagnall2017great}}. However, its time complexity of $O(n^2 \cdot l^4)$ (where $n$ is the number of time series in the dataset and $l$ is the length of a time series) makes its use in real-time settings very limited and impractical.

To overcome this challenge, machine learning based approaches have been introduced that can replace the threshold-based statistical detection~\cite{fallet2019can, schack2017computationally} . However, conventional machine learning based approaches need extraction of pre-selected features which can be labor-intensive. Recent advancements in deep learning is taking over the analysis of time-series data. For example, the deep residual network architecture ~\cite{wang2017time} can achieve the almost same accuracy of HIVE-COTE. Moreover, such deep neural network (DNN) based methods eliminate the requirement of feature engineering and selection. As a result, there have been multiple works using DNN on time-series data~\cite{nweke2018deep, che2017boosting, ismail2018evaluating, zheng2014time, liu2018time}.
DNNs on time-series data have been extensively applied to physiological signals in recent times~\cite{faust2018deep, acharya2017automated, tan2018application, wand2016deep, shen2019ambulatory, das2022bayesbeat}. Though all physiological signals are time-series data, they have different collection processes, features and applications. For example, EMG signal analysis for gesture, speech recognition~\cite{geng2016gesture, wand2016deep}, EEG signal analysis for brain computer interface~\cite{schirrmeister2017deep, acharya2018deep}, ECG signal analysis for cardiovascular diseases~\cite{acharya2017automated, tan2018application}, EOG analysis for eye movements~\cite{du2017detecting, zhu2014eog}. Among all these signals, ECG signal is the most studied one due to its least noisy nature.

However, the above existing works that used deep learning models in physiological signals are based on supervised learning and hence, suffer from data annotation problems. In the health domain, manual annotation is very time-consuming and expensive which makes it difficult to annotate a large dataset manually. 
So, obtaining a dataset with 100\% accurate ground-truth labeling is not feasible in the physiological signal dataset domain. Moreover, mislabelled training samples affecting the performance and biasing supervised classifiers is a very common issue in this domain.

To solve the above problems, in this paper, we propose a contrastive self-supervised deep learning method for physiological signal based patient similarity detection. Intuitively, contrastive learning is a representation learning framework that learns similar embeddings of positive pairs of samples (e.g., two similar samples in some sense) and ensures that the embeddings of negative pairs (e.g., two very different samples) are different from each other~\cite{stanfordcontrastivelearning}.

In our case, the signals from patients of the same disease can be considered as positive pairs and vice versa. Thus, we can use more pairs of samples in self-supervised learning (SSL) than in supervised learning -- which can mitigate the data annotation issue of physiological datasets. For example, if a dataset has $n$ and $m$ labeled signals for sick and healthy patients, respectively, a supervised learning method can use $(n+m)$ samples at most, whereas our contrastive learning based approach can use $(n \times m)$  pairs to train the model. Recent work~\cite{liu2021self} showed that SSL is more robust to dataset imbalance.

In this paper, we first present a generic self-supervised contrastive learning framework for finding physiological signal based patient similarity. To demonstrate the efficacy of our proposed framework, we applied and tested our framework in detecting \text{Atrial Fibrillation (AF)\ } since it is the most common arrhythmia~\cite{wyndham2000atrial}.  
Most of the existing approaches to detect AF are based on ECG signals~\cite{niu2019inter, acharya2017automated, xia2018detecting, yao2017atrial, yuan2016automated}. But such approaches are not applicable to prolonged monitoring with low cost. This motivated us to focus on \text{Photoplethysmography (PPG)} signals. Most of the wearables (e.g. smartwatches) nowadays are equipped with low-cost easy-to-implement optical sensors that can measure the PPG signals. As a result, it is possible to monitor patients continuously by analyzing the PPG signals. 

PPG signals have been recently used for AF detection using various approaches~\cite{schack2017computationally, chan2016diagnostic, lee2012atrial, fallet2019can} including DL techniques ~\cite{aliamiri2018deep, voisin2018ambulatory, shashikumar2018detection, kwon2019deep, gotlibovych2018end}. ~\cite{shen2019ambulatory} adapted ResNeXt architecture for 1D PPG data to detect AF. However, such a computationally heavy model might not be suitable for low-resource devices which is one of the sole reasons for using PPG signals. ~\cite{torres2020multi} proposed an unsupervised transfer learning through convolutional denoising autoencoders (CADE). But for transfer learning, the parent model they used was trained using supervised learning. Though AF detection using PPG signal is promising, it has major challenges (e.g., noise, motion artifact, intra, and inter-patient variability, etc.).~\cite{das2022bayesbeat} leveraged Bayesian deep learning to mitigate the noise issue PPG signals and provided an uncertainty estimate of the prediction. All the prior works had to use some kind of supervised learning which requires manual labeling of individual signals. To the best of our knowledge, we are the first to use self-supervised contrastive learning on PPG signals to detect AF. In summary, the main contributions of this paper are as follows: 
\begin{itemize}
    \item We propose a novel Contrastive Learning based approach, namely SimSig, for patient similarity search on physiological signal data.
    \item As a self-supervised approach, SimSig can work on a partially-labeled dataset, which overcomes a key bottleneck of labeling medical data records.
    \item Our detailed experimental study with real datasets shows that SimSig has achieved better accuracy in AF detection than the existing state-of-the-art approaches.

\end{itemize}


\section{Methods: SimSig}
\label{sec:method}
In this section, we first give a formal definition of our problem, and then discuss the key concept of contrastive learning. After that, we provide the formulation of our contrastive learning based patient similarity framework, which we named SimSig. Then we present the network architecture, the training details of SimSig and neighbor selection algorithms.



\subsection{Problem Definition}


Suppose $P = \{P_{1}^1, \dots, P_{1}^{n_1}, \dots,  P_{i}^1, \dots, P_{i}^{n_i}, \dots, P_{N}^1, \dots, \\P_{N}^{n_N} \}$ for $i = 1,2,...,N$ is a patient database of $N$ individuals, each having multiple time series signal segments obtained from sensors where patient $i$ has $n_i$ segments. Each segment $P_{i}^j$ for $j = 1,2,...,n_i$ is of length $l$, i.e., $P_{i}^j\in \mathbb{R}^l$. Our goal is to predict label $y \in \{0,1\}$ for a query individual $Q$ based on the labels from the patient database.  



\subsection{Contrastive Representation Learning}
Contrastive representation learning or in short, Contrastive learning is a popular form of self-supervised learning that encourages augmentations of the same input to have more similar representations or embeddings compared to augmentations or embeddings from different inputs. The key idea of contrastive representation learning is to contrast semantically similar and dissimilar pairs of data points to make the representation of similar pairs closer, and those of dissimilar pairs more orthogonal by minimizing the contrastive loss~\cite{van2018representation, chen2020simple}.

\subsection{Model Architecture of SimSig}

Inspired by the framework proposed by~\cite{chen2020simple} named SimCLR, we design our network adopting some of the core components of SimCLR. The SimCLR learns representation by maximizing the agreements between different augmentations of similar categories of examples. 
Since we intend to learn the similarity within the segments (portions) of physiological signals of the same type of patients, we maximize agreements between segments from the same individual and examine how this helps to learn similarities across patients with similar physiological signals. Our adopted framework comprises the following three major components as demonstrated in Figure~\ref{fig:sim_model_archi}:
\begin{itemize}
    \item A neural network based encoder $f(.)$ to extract representation vectors from signal segments. We pass input vector $P_i^k$ to obtain representation vector $h_i^k = f(P_i^k)$. {In our case, we use 1D ResNext50 as our encoder architecture.} 
    \item A small neural network projection head $g(.)$ similar to~\cite{chen2020simple} that maps the representation vectors generated by the encoder on which contrastive loss is applied. An MLP with one hidden layer is used to obtain $z_i^k = g(h_i^k)$
    \item A contrastive loss function to apply on $z_i^k$'s. We consider custom defined contrastive loss functions below.
\end{itemize}

\subsubsection*{Contrastive Loss Functions}
\label{subsec:loss}
We consider two contrastive loss functions: \textit{NT-Xent} loss as defined in~\cite{chen2020simple} and another variant of it proposed by us. We name the latter one \textit{NT-Xent Multi} loss.

In the NT-Xent loss shown in Equation~\ref{eqn:loss_fn_ntxent_indiv}, the same sample is augmented to produce a pair, and the model is trained to increase the similarity between the pair. In our case, we only sample two segments from the same individual in a batch and consider them as a pair.

\begin{equation}
    l_\alpha = -log\frac{exp(sim(z_\alpha, z_\beta)/\tau)}{
    \sum_{\gamma=1}^{N} [1_{[\gamma \notin \{\alpha, \beta\}]} exp(sim(z_\alpha, z_\gamma)/\tau)]
    }
 \label{eqn:loss_fn_ntxent_indiv}
\end{equation}
Here, $\beta$ is a segment such that $\beta \neq \alpha$ and $ind(\beta)=ind(\alpha)$. $\tau$ is a hyper-parameter. {Note that the similarity function (cosine similarity in our case) has been applied on the output of the projection head i.e. $z$'s.} Finally,  the total loss is aggregated as:
\begin{equation}
    L = \sum_{i=1}^{N} l_i
    \label{eqn:loss_fn_ntxent}
\end{equation}

However, this imposes a limit on the batch size. Hence, we designed another loss function, NT-Xent Multi loss.
For the NT-Xent Multi loss, we randomly sample a minibatch of $K$ examples. In these $K$ examples, say there are examples from $N_K$ unique individuals ($N_K \leq K$) and the loss function for $\alpha$-th sample is defined as:

\begin{equation}
    l_\alpha = -log\frac{\sum_{\beta=1}^{K} [1_{[ind(\beta)=ind(\alpha)]} exp(sim(z_\alpha, z_\beta)/\tau)]}{\sum_{\gamma=1}^{K} [1_{[ind(\gamma) \neq ind(\alpha)]} exp(sim(z_\alpha, z_\gamma)/\tau)]}
\end{equation}

Here, $\tau$ is a hyper-parameter. {In summary, the numerator part inside the logarithm function is the summation of exponential similarities between segments from the same individual, which we consider as positive samples. The denominator part is the summation of exponential similarities between segments from different individuals which we consider as negative samples.}
Finally, the total loss function is defined like the previous one as:
\begin{equation}
    L = \sum_{i=1}^{N} l_i
    \label{eqn:loss_fn}
\end{equation}



\begin{figure*}
    \centering
    \includegraphics[width=\textwidth]{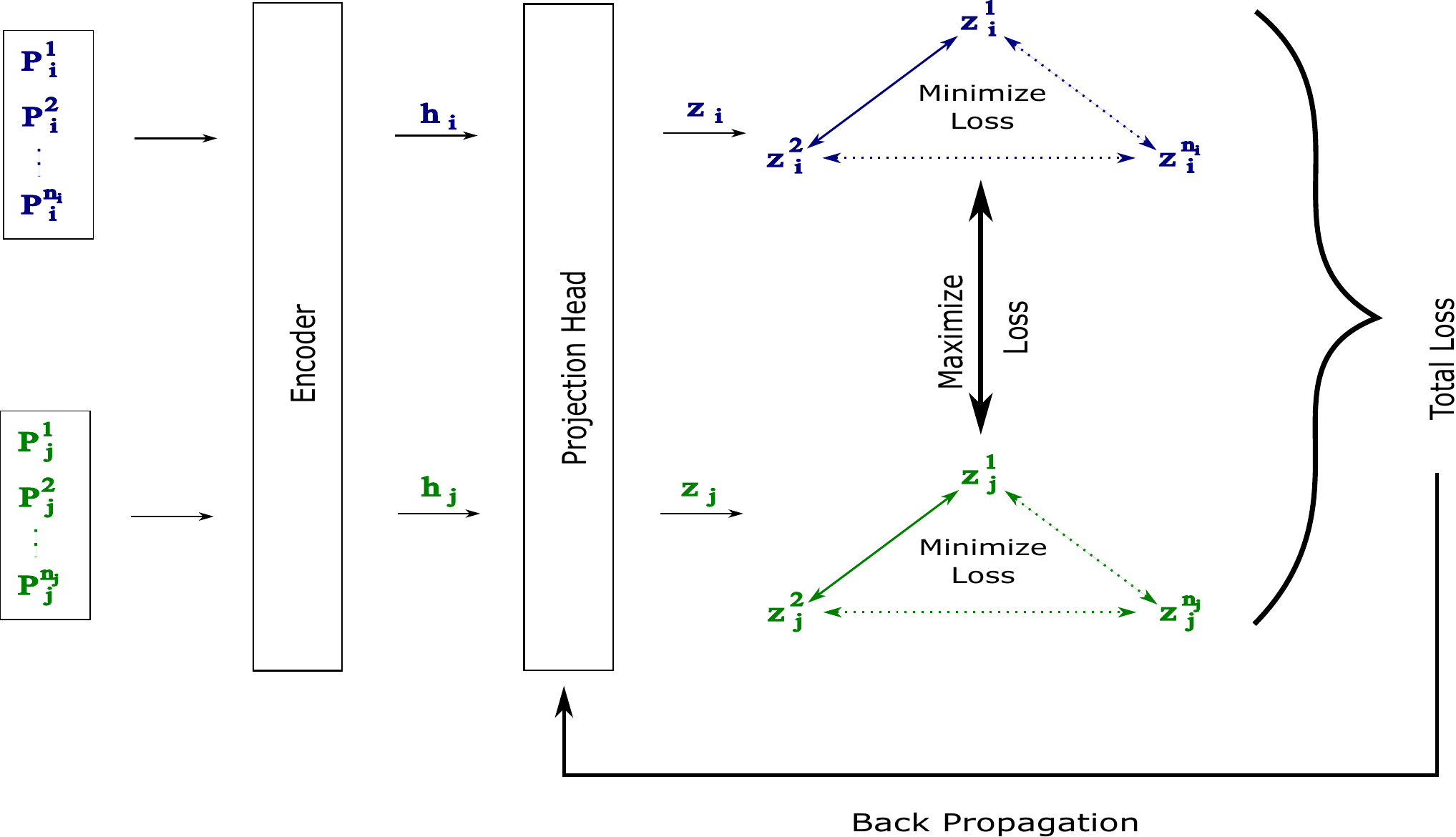}
    \caption{SimSig Contrastive Representation Learning Training Pipeline}
    \label{fig:sim_model_archi}
\end{figure*}

The similarity learning network architecture is demonstrated in Figure~\ref{fig:sim_model_archi}. 
The encoder network, 1D ResNext50 (in this case), takes 1D signal segments and generates the corresponding embeddings of size 1024.
We denote the embedding of $k$-th segment of the $i$-th individual as $h_i^k$. The embeddings: $h_i^k$'s are then passed to the projection layer to generate $z_i^k$'s. The projection layer consists of a linear layer followed by a ReLU activation, then another linear layer to generate $z_i^k$. 

\subsection{Inference Phase}

\begin{figure*}
    \centering
    \includegraphics[scale = 0.45]{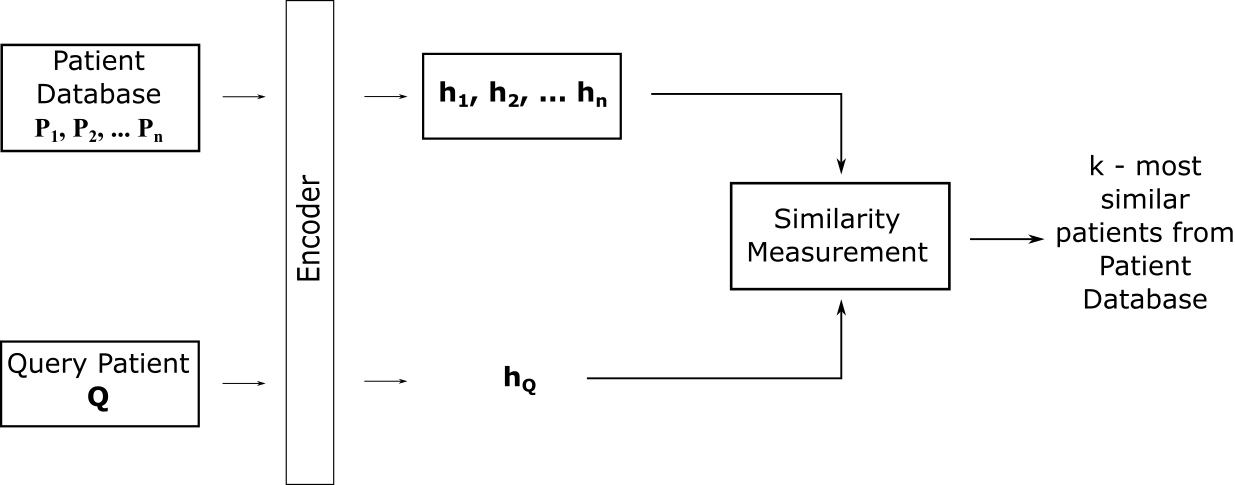}
    \caption{Patient Similarity Detection using Our Model}
    \label{fig:patient_sim_archi}
\end{figure*}

In this phase, our framework works in two major steps as demonstrated in Figure~\ref{fig:patient_sim_archi}. First, it generates the embeddings of signal segments from an individual using the encoder model we trained earlier using contrastive representation learning. After that, we perform a number of similarity measurements and neighbor selections with the embeddings of individuals we have in our database. Based on the neighbors, we label a new individual.

Once the network is trained by minimizing the contrastive loss, we generate the embeddings of training samples to be later used for patient similarity. We refer to this as \textbf{\textit{Patient Database}}. Note that in order to achieve the best performance according to~\cite{chen2020simple}, we use $h_i^j$'s for the embeddings. We similarly obtain embedding $q_i^j$'s for time series segments from a query patient, $Q$ whose label we want to infer.

\subsection{Neighbor Selection}

From the \emph{SimSig Encoder} model, we get an embedding vector for each segment. We keep the embeddings of individuals from the training set that we refer to as the Patient Database. 
During the evaluation, we apply different metrics for this distance calculation between a query individual and an individual from the Patient Database. Figure~\ref{fig:patient_sim_archi} represents the generic model of distance calculation. 
In Figure \ref{fig:patient_sim_archi}, the $P_i$ represents the patients with id $i$. The $i$-th patient has $n_i$ number of segments of PPG signal, and each segment is represented by ${P}_i^j$, where $i$ is the patient id and $j$, is the $j$-th segment of a signal. After feeding these segments to our encoder, we get the embedding vector $h_i^j$ for each ${P}_i^j$. 

To calculate the pair-wise distance between a query patient and a patient from the database, {we first calculate the distance between every pair of embedding vectors where one is from a query patient and the other one is from a patient in the patient database}. 
{For example, to calculate the distance (or the similarity in our context) from query patient $Q$ to the $i$-th patient $P_i$ in our patient database, the $h_Q^j$ embedding vectors are multiplied with the embeddings of the $i$-th patient i.e. $h_i^k$'s in the patient database.}

After that, we used different metrics to calculate the distance between the two individuals 
We denote $d_i$ as the final distance between the query patient and $i$-th patient where $i$ represents the id of the patient in the patient database with whom the distance is to be calculated. Thus, for a query patient, we calculate the final distance $d_i$'s for each patient in our Patient Database where $i=[1, N]$, $N = $~number of individuals in the database.
Consider $P_i$ to be an individual in database with $n_i$ signal embeddings $h_{i}^1, h_{i}^2, ...., h_{i}^{n_i}$. Let $Q$ be a test individual with $M$ embeddings $h_{Q}^{1}, h_{Q}^{2}, ...., h_{Q}^{M}$. We denote $d_{i}^{j, m}$ to be the distance between $h_{i}^j$ and $h_{Q}^{m}$. We experimented with the following distance calculation criteria to measure the distance between the signals of two individuals:

\subsubsubsection{Overall Min Distance} 
Here we consider the minimum distance between two segments where each one is from a separate individual as the distance between those two individuals.
We first calculate the pairwise cosine distances between every pair of segments between $P_i$ and Q. We calculate $n_i \times m$ cosine distances and finally we take distance $d_{i}$ to be:
\begin{equation*}
d_{i} = min(d_{i}^{1, 1}, d_{i}^{1, 2}, \dots, d_{i}^{j, m},\dots, d_{i}^{n_i, m})   
\end{equation*}
as the distance between the query individual Q and $P_i$ with respect to their signal similarity. We get $N$ such distances for the whole database. Finally, we choose the $k$ nearest individuals to $Q$ in terms of distance. We label $Q$ as AF if the majority of the $k$ neighbors are AF individuals, otherwise Non-AF.
\\
\subsubsubsection{Average Min Distance} 
In this case we consider the average of all the distances between all possible pairs of segments where each one is from a separate individual as the distance between those two individuals.
We calculate the $d_{i}^{j, m}$'s similarly to \textit{Overall Min Distance} criteria. Then finally we take,
\begin{equation*}
d_{i} = average(d_{i}^{1, 1}, d_{i}^{1, 2}, \dots, d_{i}^{j, m},\dots, d_{i}^{n_i, m})   
\end{equation*}
to be the distance between the query individual Q and $P_i$ with respect to their signal similarity. Finally, we label $Q$ as previously discussed.

\subsubsubsection{Weighted Average Min Distance}
We also consider a weighted version of the average min distance by weighting the individual label 
with $1/{d_{i}^{j, m}}^2$ i.e. the inverse of the squared distance.

\subsubsubsection{Pct Min Distance} 
In this setting for an individual, we calculate all possible pairs of distances with every segment from all other neighbors i.e. individuals in the database. Among the neighbors we consider the closest neighbors to be the individuals who have the most number of segments with distances that are below a certain value, which we call the \emph{radius}.
We first choose a hyper-parameter, radius, $r$. Then we calculate a count for every individual in the database. The count for individual $i$ in the database is defined as:
\begin{equation*}
    \resizebox{0.48\textwidth}{!}{$C_{i} = \frac{|\{j: \text{cosine distance}(h_i^j, h_{Q}^m) \leq r \text{ for } j=[1, n_i] \text{ and }m=[1, M]\}|}{M}$}
\end{equation*}

We take the top $k$ individuals with the highest $C_{i}$ values. We infer the label of $Q$ similarly to the aforementioned ones.



\section{Experiments}
\label{sec:experiment}
In our experiment, we have selected the Atrial Fibrillation (AF) detection problem to be our case study. In this section, we first give the description of the AF detection problem followed by the description of the dataset that we use as an example to apply our framework. After that we provide the implementation details and training environment configurations for two Simsig versions. Then we define some of the evaluation metrics we use in our experiment. We have run extensive experiments with all the configurations by varying the hyper-parameters, and report them in the Result section.

\subsection{AF Detection Problem}

Given a set of PPG signal segments of an individual, we want to detect whether the person has atrial fibrillation (AF) or not. Atrial Fibrillation (AF) is a type of abnormality characterized by irregular beating of the two upper chambers of the heart. 

For our case-study, we will detect negative (Non-AF) or positive (Atrial Fibrillation) for the individual whose signal segments from $Q$ and our time series data consists of PPG signals obtained from wearable sensors.


\subsection{Dataset}

We have used the largest publicly available dataset, which we refer to as the \textit{Stanford Wearable Photoplethysmography Dataset}\footnote{https://www.synapse.org/\#!Synapse:syn21985690/files/} for training our model and evaluating with the state-of-the-art. ~\cite{torres2020multi} made this dataset public with their work \emph{DeepBeat}. 
The dataset contains signal segments collected using wrist-worn wearable devices. There are more than 500K segments from a total of {175 individuals (108  AF subjects and 67 non-AF subjects)}, each with duration of 25s sampled at 128 Hz and later downsampled to 32 Hz. The starting timestamp of each segment is also provided along with the dataset. 

The dataset includes three categories of signal labels, labeled as Poor, Good, and Excellent. However, only a portion of these labels were assigned by humans; the remaining labels were generated by a model trained on the human-labeled portion, resulting in imprecise labels in the dataset. This can cause downstream models to be susceptible to error propagation, as demonstrated in~\cite{das2022bayesbeat}.

\begin{table*}[hbt!]
\begin{center}
\caption{\label{tab:dataset_table_bayesbeat} Distribution of the Revised Dataset from~\cite{das2022bayesbeat}}
\begin{tabulary}{\linewidth}{|C|C|C|C|C|C|C|C|C|C|C|C|C}     
    \hline
    \textbf{Set} & \textbf{\#Individuals} & \textbf{\#Samples} & \textbf{\#AF Samples} & \textbf{\#Non-AF Samples} & \textbf{\#AF Individuals} & \textbf{\#Non-AF Individuals} & \textbf{Size ratio} \\ 
    \hline
    Train & 132 & 108171 & 41511 & 66660 & 82 & 50 & 0.698 \\
    \hline
    Validation & 20 & 22294 & 8310 & 13984 & 12 & 8 & 0.144 \\
    \hline
    Test & 23 & 24579 & 9703 & 14876 & 14 & 9 & 0.159 \\
    \hline

\end{tabulary}
\end{center}
\end{table*}

The provided dataset also has some distribution issues present in the original train, validation, test split. This distribution issue has been addressed and a proper redistribution has been provided by~\cite{das2022bayesbeat}. Hence, we adapt the distribution from~\cite{das2022bayesbeat} with a split of 70\% as train, 15\% as validation, and 15\% as test sets where no subject is shared among different sets and all overlapping signal segments from the validation and test sets are removed using the description of the split provided by~\cite{das2022bayesbeat}. The distribution is shown in Table~\ref{tab:dataset_table_bayesbeat}

\subsection{Implementation Details}
The cost function in Equation~\ref{eqn:loss_fn} is  optimized by minibatch gradient descent. We have used the Adam optimizer variant since it gives the best performance in our experiments. We sample 512 segments for a minibatch that might contain multiple samples from the same individual and minimize the total loss value according to it. Figure~\ref{fig:sim_model_archi} represents the training pipeline for similarity learning. Note that the loss values among $z_i^j$'s of similar individuals are minimized while loss with other individuals is maximized. 

We implement the training pipeline for training our model in PyTorch~\cite{paszke2019pytorch}. 
We selected $batch\_size=512$ for NT-Xent Multi loss configuration and trained for 50 epochs with a learning rate of $1x10^{-3}$ with Adam~\cite{kingma2014adam} as the optimizer using an i7-7700 workstation with 32GB of RAM, and a GTX 1070 GPU for a day.

\subsection{Evaluation Metrics and Baselines}
We have employed a diverse set of metrics, such as Recall (Sensitivity), Specificity (True Negative Rate, TNR), Precision (Positive Predictive Value, PPV), F1-score, and Accuracy, to evaluate various models. The formal expressions of these metrics are as follows.
    


\begin{align*}
    Recall &= \frac{TP}{TP + FN} \\
    Specificity &= \frac{TN}{TN + FP} \\
    Precision &= \frac{TP}{TP + FP} \\
    \\
    F1 &= \frac{2*(Precision*Recall)}{Precision + Recall} \\
    \\
    Accuracy &= \frac{TP + TN}{TP + TN + FP + FN}
\end{align*}



\vspace{1mm}

In this context, TP (True Positive) refers to the number of positive individuals (AF patients) that the model correctly identified, TN (True Negative) refers to the number of negative individuals (non-AF patients) that the model correctly identified, FP (False Positive) refers to the number of individuals that the model incorrectly classified as positive, and FN (False Negative) refers to the number of individuals that the model incorrectly classified as negative.



Most of the works in this field have been limited to segment wise prediction whereas our work is on individual-level. 
\cite{shen2019ambulatory, torres2020multi} have been the state-of-the-art works for predicting AF/Non-AF for a PPG segment. We adapted their approach to predict AF/Non-AF for an individual. For an individual, if more than half of the segments results into AF predictions for these models, we label the individual as AF.

\section{Results}

In this section, we present the results of our experiments. First, we analyze SimSig for various neighbor sizes and configuration parameters.

\subsection{Model performance and Ablation Study}

First, we present the results on the validation set for the configurations with the topmost performances in Table~\ref{tab:Perf_table_valid} with five different sub-tables separately for different values of the neighbor size $k$. 
We have set the value of hyper-parameter, $\tau=0.5$ for both loss functions NT-Xent and NT-Xent Multi loss.

\begin{table*}[h]
\caption{\label{tab:Perf_table_valid}Performance of SimSig on Validation for different configurations}
	\begin{subtable}[h]{\textwidth}
        \begin{center}
        \caption{\label{tab:Perf_table_valid_k3}For k=3}
        \scalebox{0.9}{
        \begin{tabular}{llccccc}
            \toprule
            Model Loss Function & Configuration & Sensitivity & Specificity & Precision & F1 & Accuracy\\
            \midrule
            \multirow{3}{*}{NT-Xent} 
            & Average Min Distance & 0.167 & 1 & 1& 0.286 & 0.5\\
            & Overall Min Distance & 0.667 & 0.75 & 0.8 & 0.727 & 0.7\\
            & Pct Min Distance & 0.333 & 1 & 1 & 0.5 & 0.6\\
            \midrule
            \multirow{3}{*}{NT-Xent Multi} 
            & Average Min Distance & 0.917 & 1 & 1 & 0.957 & 0.95\\
            & Overall Min Distance & 0.917 & 0.75 & 0.846 & 0.88 & 0.85\\
            & Pct Min Distance & 0.5 & 1 & 1 & 0.667 & 0.7\\
            \bottomrule
        \end{tabular}
        }
        \end{center}
	\end{subtable}
	\hfill
        \linebreak

	\begin{subtable}[h]{\textwidth}
        \begin{center}
        \caption{\label{tab:Perf_table_valid_k5}For k=5}
        \scalebox{0.9}{
        \begin{tabular}{llccccc}
            \toprule
            Model Loss Function & Configuration & Sensitivity & Specificity & Precision & F1 & Accuracy\\
            \midrule
            \multirow{3}{*}{NT-Xent} 
            & Average Min Distance & 0.167 & 1 & 1& 0.286 & 0.5\\
            & Overall Min Distance & 0.667 & 0.75 & 0.8 & 0.727 & 0.7\\
            & Pct Min Distance & 0.333 & 1 & 1 & 0.5 & 0.6\\
            \midrule
            \multirow{3}{*}{NT-Xent Multi} 
            & Average Min Distance & 0.917 & 1 & 1 & 0.957 & 0.95\\
            & Overall Min Distance & 0.833 & 0.75 & 0.833 & 0.833 & 0.8\\
            & Pct Min Distance & 0.5 & 1 & 1 & 0.667 & 0.7\\
            \bottomrule
        \end{tabular}
        }
        \end{center}
	\end{subtable}
	\hfill
        \linebreak

        \begin{subtable}[h]{\textwidth}
        \begin{center}
        \caption{\label{tab:Perf_table_valid_k7}For k=7}
        \scalebox{0.9}{
        \begin{tabular}{llccccc}
            \toprule
            Model Loss Function & Configuration & Sensitivity & Specificity & Precision & F1 & Accuracy\\
            \midrule
            \multirow{3}{*}{NT-Xent} 
            & Average Min Distance & 0.083 & 1 & 1 & 0.154 & 0.45\\
            & Overall Min Distance & 0.667 & 0.75 & 0.8 & 0.727 & 0.7\\
            & Pct Min Distance & 0.333 & 1 & 1 & 0.5 & 0.6\\
            \midrule
            \multirow{3}{*}{NT-Xent Multi} 
            & Average Min Distance & 0.917 & 1 & 1 & 0.957 & 0.95\\
            & Overall Min Distance & 0.833 & 0.75 & 0.833 & 0.833 & 0.8\\
            & Pct Min Distance & 0.5 & 1 & 1 & 0.667 & 0.7\\
            \bottomrule
        \end{tabular}
        }
        \end{center}
	\end{subtable}
	\hfill
        \linebreak

	\begin{subtable}[h]{\textwidth}
        \begin{center}
        \caption{\label{tab:Perf_table_valid_k9}For k=9}
        \scalebox{0.9}{
        \begin{tabular}{llccccc}
        \toprule
            Model Loss Function & Configuration & Sensitivity & Specificity & Precision & F1 & Accuracy\\
            \midrule
            \multirow{3}{*}{NT-Xent} 
            & Average Min Distance & 0.083 & 1 & 1 & 0.154 & 0.45\\
            & Overall Min Distance & 0.667 & 0.75 & 0.8 & 0.727 & 0.7\\
            & Pct Min Distance & 0.333 & 1 & 1 & 0.5 & 0.6\\
            \midrule
            \multirow{3}{*}{NT-Xent Multi} 
            & Average Min Distance & 0.833 & 1 & 1 & 0.909 & 0.9\\
            & Overall Min Distance & 0.75 & 0.75 & 0.818 & 0.783 & 0.75\\
            & Pct Min Distance & 0.583 & 1 & 1 & 0.737 & 0.75\\
            \bottomrule
        \end{tabular}
        }
        \end{center}
	\end{subtable}
	\hfill
        \linebreak

	\begin{subtable}[h]{\textwidth}
        \begin{center}
        \caption{\label{tab:Perf_table_valid_k11}For k=11}
        \scalebox{0.9}{        
        \begin{tabular}{llccccc}
            \toprule
            Model Loss Function & Configuration & Sensitivity & Specificity & Precision & F1 & Accuracy\\
            \midrule
            \multirow{3}{*}{NT-Xent} 
            & Average Min Distance & 0.167 & 1 & 1 & 0.286 & 0.5\\
            & Overall Min Distance & 0.75 & 0.75 & 0.818 & 0.783 & 0.75\\
            & Pct Min Distance & 0.333 & 1 & 1 & 0.5 & 0.6\\
            \midrule
            \multirow{3}{*}{NT-Xent Multi} 
            & Average Min Distance & 0.833 & 1 & 1 & 0.909 & 0.9\\
            & Overall Min Distance & 0.75 & 0.75 & 0.818 & 0.783 & 0.75\\
            & Pct Min Distance & 0.583 & 1 & 1 & 0.737 & 0.75\\
            \bottomrule
        \end{tabular}
        }
        \end{center}
	\end{subtable}
\end{table*}

We have tried each configuration by varying neighbor size, $k$ to be 3, 5, 7, 9, and 11, using both loss functions  NT-Xent and NT-Xent Multi. 
The table ~\ref{tab:Perf_table_valid} shows performances separately for all the distance metrics we defined earlier for each configuration. However, weighted versions of Average Min and Overall Min performed exactly the same as the unweighted versions. 
Hence, we did not report their performance separately.

For the network using NT-Xent loss, we can observe from Table~\ref{tab:Perf_table_valid} that the neighbor selection criteria `Overall Min Distance' performs the best for all of $k=3, 5, 7, 9~and~11$. While for the network using NT-Xent Multi loss, the neighbor selection criteria `Average Min Distance' shows the best performance for different $k$ values. Hence, we select the `Overall Min Distance' metric for the network using NT-Xent loss and `Average Min Distance' metric for the network using NT-Xent Multi loss.

\begin{figure}
    \centering
    \includegraphics[scale = 0.45]{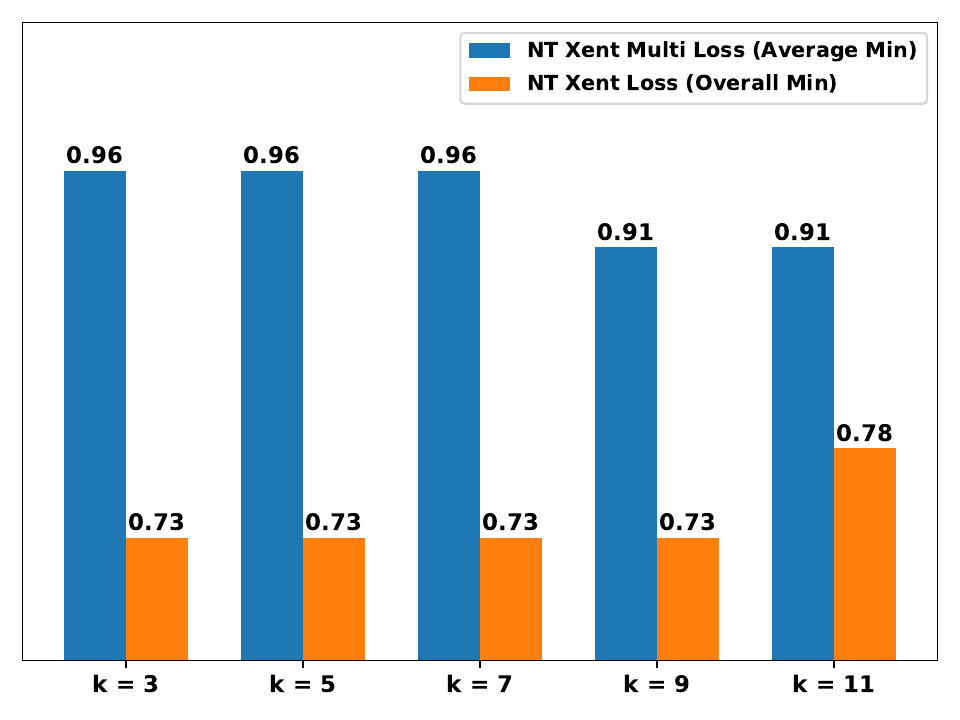}
    \caption{Selection of best neighbor size k}
    \label{fig:best_k_barplot}
\end{figure}


Figure~\ref{fig:best_k_barplot} represents the best F1 score for both losses considering the best configuration for each of them. We can observe that NT-Xent Multi loss with `Average Min Distance' outperforms NT-Xent loss with `Overall Min Distance' for every value of $k$. Also for $k=3, 5, 7$, NT-Xent Multi loss with `Average Min Distance` provides a similar performance of $0.96$ F1 score which drops down to $0.91$ with higher $k$ values. Therefore, we choose $k=7$ for the `Average Min Distance' metric since it is expected to have a better prediction and more confidence to practitioners because of taking a decision from a higher number of neighbors from the Patient Database. Hence, we choose NT-Xent Multi loss as our SimSig loss function with `Average Min Distance' as the neighbor selection metric with neighbor size $k=7$ to compare it with other baseline methods in the next subsection.

\subsection{Comparison with baseline methods}
We compare the performance of the SimSig on the test set with two other baseline methods, namely ResNext~\cite{shen2019ambulatory} and DeepBeat~\cite{torres2020multi}, that have been adapted to work with individuals. Since these two baselines predict on signal segments, in order to generate individual-wise AF/Non-AF labels we considered an individual as AF if he/she receives more AF labeled signal segments than Non-AF labeled signal segments by the model. We report the results in Table~\ref{tab:baseline_stanford}.

\begin{table*}[htb]
\begin{center}
\caption{\label{tab:baseline_stanford}Comparison of Simsig with other baseline models on Revised Stanford Wearable Photoplethysmography Dataset}
\begin{tabular}{lccccc}
    \toprule
     & Sensitivity & Specificity & Precision & F1 & Accuracy \\
    \midrule
    \cite{shen2019ambulatory} ResNext & 0.714 & 0.889 & 0.909 & 0.8 & 0.783 \\
    \cite{torres2020multi} (\emph{Deepbeat}) & 0.857 & 0.778 & 0.857 & 0.857 & 0.826\\
    \cite{torres2020multi} (\emph{DeepBeat (Excellent)}) & 0.6 & 0.714 & 0.75 & 0.667 & 0.647\\
    \cite{torres2020multi} (\emph{DeepBeat (Non-Poor)}) & 0.75 & 0.714 & 0.818 & 0.783 & 0.737\\
    Simsig (NT-Xent Multi loss, Average Min Distance, $k=7$) & 0.929 & 0.667 & 0.813 & 0.867 & 0.826\\
    \bottomrule
\end{tabular}
\end{center}
\end{table*}

Table~\ref{tab:baseline_stanford} demonstrates the performance of different methods considering sensitivity (recall), specificity, precision, F1-score, and accuracy of SimSig with those of other baseline methods on the test set of the dataset. We present the performance of SimSig with the configuration of `Average Min Distance' ($k=7$) with NT-Xent Multi loss since it yields the best performance on the validation set.
We also show the performance metrics of DeepBeat~\cite{torres2020multi} on three settings based on their signal quality prediction. 

We observe that SimSig with the mentioned configuration outperforms other baseline models for overall metrics like F1-score and accuracy. Compared to~\cite{shen2019ambulatory}, SimSig has 6.7\% higher F1-score and 4.3\% higher accuracy. When compared to Deepbeat~\cite{torres2020multi}, SimSig outperforms it by 1-20\% for F1-score and up to 17.9\% for accuracy depending on its settings.






\section{Conclusion}
\label{sec:conclusion}
In this paper, we proposed a novel framework to learn the similarity between patients from their physiological signals using \textbf{self-supervised} contrastive learning and neighbor selection. Our main focus was to address the \textbf{data annotation} issue that causes the supervised approaches to suffer. As a case study, we have selected the Atrial Fibrillation detection problem from the photoplethysmography signal. We have thoroughly experimented with our framework on the dataset for our case study varying several hyper-parameters: neighbor selection criteria, neighbor size, etc. From the comparison with other baseline methods, we find that our framework for finding patient similarity performed substantially better than those.


\section{Code Availability}
\label{sec:codebase}
Trained weights and relevant source codes of \emph{SimSig} are publicly available at this github link:\\
\url{https://github.com/Subangkar/Simsig}





\bibliographystyle{unsrt}

\bibliography{bibilography}



\end{document}